\documentclass[twocolumn,prl,preprintnumbers,showpacs,amsmath,amssymb]{revtex4}
\usepackage{graphicx}
\usepackage{dcolumn}
\usepackage{bm}
\begin{document}

\title [] {Structure, bonding, and hardness of CrB$_4$: a superhard material?}

\author{Haiyang Niu$^1$}
\author{Jiaqi Wang$^1$}
\author{Xing-Qiu Chen$^1$}
\email[Corresponding author: ]{xingqiu.chen@imr.ac.cn}
\author{Dianzhong Li$^1$}
\author{Yiyi Li$^1$}
\author{Petr Lazar$^{2,3}$}
\author{Raimund Podloucky$^2$}
\author{Aleksey N. Kolmogorov$^4$}

\affiliation{$^1$ Shenyang National Laboratory for Materials
Science, Institute of Metal Research, Chinese Academy of Sciences,
Shenyang 110016, China}

\affiliation{$^2$ Center for Computational Materials Science,
University of Vienna, Sensengasse 8, A-1090 Vienna, Austria }

\affiliation{$^3$ Regional Centre of Advanced Technologies and
Materials, Department of Physical Chemistry, Faculty of Science,
Palacky University Olomouc, Tr. 17. Listopadu 12, 771 46 Olomouc,
Czech Republic}

\affiliation{$^4$ Department of Materials, University of Oxford,
Parks Road, Oxford OX1 3PH, United Kingdom}

\date{\today}

\begin{abstract}
By electron and X-ray diffraction we establish that the CrB$_4$
compound discovered over 50 years ago crystallizes in the $oP10$
(\emph{Pnnm}) structure, in disagreement with previous experiments
but in agreement with a recent first-principles prediction. The 3D
boron network in the new structure is a distorted version of the
rigid carbon $sp^3$ network proposed recently for the high-pressure
C$_4$ allotrope. According to our density functional theory
calculations and the analysis of the bonding, CrB$_4$ is a
potentially superhard material. In fact, the calculated weakest
shear and tensile stresses exceed 50 GPa and its Vickers hardness is
estimated to be 48 GPa.
\end{abstract}

\pacs{62.40.+i, 62.20.Qp, 71.20.Be, 63.20.dk}

\maketitle

Covalent networks with high atomic densities and three-dimensional
(3D) morphologies \cite{1,1-1,1-2,1-3,1-4} are basic features of
most of the known superhard materials, including diamond [Fig.
1(a)], \emph{c}-BC$_2$N, \emph{c}-BN, and the recently found
compounds \emph{c}-BC$_5$ \cite{2} and $\gamma$-B$_{28}$
\cite{3,3-1,3-2}. Three new promising superhard allotropes of carbon
with strong quasi-\emph{sp}$^3$ covalent bonding as realized in a
monoclinic (M-carbon\cite{4}), tetragonal body-centered (bct-C$_4$
\cite{5,5-1}), and orthorhombic (W-carbon \cite{6}) structure have
been proposed for the interpretation of the X-ray diffraction
pattern of cold-compressed graphite\cite{7}. In particular,
metastable bct-C$_4$ is built up by an unusual framework \cite{7-2} of
interconnected square C$_4$ units [Fig. 1(b)] and has been predicted
to be superhard by several first-principles studies
\cite{8,8-1,8-2,8-3}. Inspired by the search for superhard materials
which can be fabricated without the need of an expensive high
pressure \cite{11} or a chemical vapor deposition \cite{Mcmillan}
methods, we re-examine a known stable intermetallic CrB$_4$ compound
comprised of similar B$_4$ units.
We find that, compared to the ReB$_2$ compound shown recently to
have a remarkably high hardness \cite{11,ReB2,ReB2-1,ReB2-2},
CrB$_4$ holds the promise to have even more outstanding mechanical
properties.

First we characterize CrB$_4$ experimentally by means of electron
diffraction (ED) and X-ray diffraction (XRD) techniques confirming a
first-principles prediction \cite{FeB4,14} that the orthorhombic
structure of CrB$_4$, originally suggested to have the $Immm$ space
group (with the $oI$10 unit cell in Pearson notation, see Fig.
1(c)), has a lower-symmetry $Pnnm$ space group ($oP$10, see Fig.
1(d)). On the basis of density functional theory (DFT) calculations
\cite{PBE,PAW,VASP,VASP-1}, we establish that CrB$_4$ with the newly
claimed structure has lowest ideal tensile and shear strengths of 51
GPa, which are comparable to those of cubic boron nitride (c-BN).
Making use of an empirical model \cite{22,22-1} correlating the
elastic moduli and Vickers hardness ($H_v$), we estimate $H_v
\approx $ 48 GPa, which exceeds significantly the 40 GPa threshold
of superhardness. To rationalize this finding we perform a DFT study
of the ten transition metal borides, \emph{TM}B$_4$ (with \emph{TM}
= Ti, V, Cr, Mn,Fe as well as \emph{TM} = Zr, Nb, Mo, Tc, Ru). Our
results indicate that the atomic size and valence of the \emph{TM}
elements play a key role in determining the mechanical properties.
The hardness reaches a maximum for TM=Cr when all bonding B
quasi-$sp^3$ and hybridized Cr-B states become occupied.

\begin{figure}[b!]
\begin{center}
\includegraphics[width=0.5\textwidth]{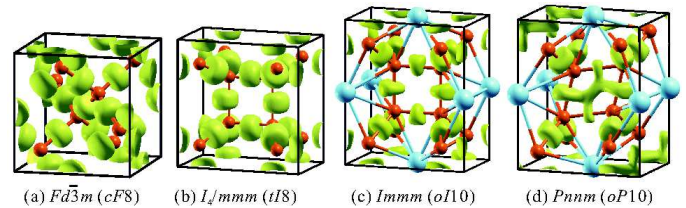}
\end{center}
\caption{(color online)  Isosurfaces of the electron localization
function (ELF) \cite{27}, corresponding to a value of 0.75. (a,b):
diamond and bct-C$_4$ carbon; (c,d): orthorhombic structures of
CrB$_4$. Small and large balls denote B and Cr atoms, respectively.
}\label{fig1}
\end{figure}

In 1968, Andersson and Lundstr\"om \cite{12} reported the synthesis
of CrB$_4$ and characterized it as an orthorhombic \emph{oI}10
structure.  Given the very good fit of the X-ray pattern to the
\emph{oI}10 structure and the recently demonstrated elastic
stability of the compound\cite{13} there was no reason to suspect
incompleteness of this structural model. However, examination of the
ground states of Fe-B \cite{FeB4} and Cr-B systems \cite{14}
revealed a dynamical instability of the oI10 structure due to phonon
modes with imaginary frequency near $\mathbf{q}=\mathbf{0}$. As a
consequence, the boron framework undergoes a significant distortion
transforming the orthorhombic body-centered structure ($oI$10) into
a primitive one ($oP$10). It was observed \cite{14} that this
structural transformation leaves the unit cell dimensions and the
XRD patterns essentially unchanged (see Fig. S2 in Ref.
\cite{RefAdd}) which necessitates the use of an alternative
characterization technique to finally resolve the structure of
CrB$_4$.


A 20g sample with the initial composition of CrB$_4$ was prepared by
repeated arc-melting of electrolytic chromium (from Alfa Aesar,
claimed purity 99.997\%) and crystalline boron pieces (from Alfa
Aesar, claimed purity 99.5\%) under argon atmosphere. Cut sample
pieces were sealed in quartz under argon and annealed in a high
temperature furnace for 192 hours at 1250 $^\circ$C. The annealed
samples were characterized via metallographic microscope (LEISS
Axiovert 200 MAT), scanning electron microscope (SEM, HITACHI
S-3400N) in the back-scattered electron mode (BSE). Our electron
probe microanalyser (EPMA, SHIMADZU EPMA-1610) results showed 20.374
at.\% and 79.626 at.\% elemental compositions of Cr and B,
respectively. The presence, distribution, and phase characteristics
of CrB$_4$ (78.06\%), CrB$_2$ (7.32\%), and amorphous boron
(14.62\%) were further analyzed with an electron backscatter
diffraction (EBSD) micrograph \cite{RefAdd}. TEM characterization of
finely ground samples\cite{RefAdd} was carried out with Tecnai G2
F20 S-TWIN transmission electron microscope. Finally, we obtained
X-ray diffraction (XRD) patterns using a Rigaku diffractometer and
Cu \emph{K}$_{a}$ irradiation ($\lambda$=1.54056 \AA\,) and
performed full Rietveld refinement using the FULLPROF package
\cite{fullprof}.

For DFT calculations we used the Perdew-Burke-Ernzerhof
exchange-correlation functional \cite{PBE} within the generalized
gradient approximation and the projector-augmented waves method
\cite{PAW} as implemented in {\small VASP}\cite{VASP,VASP-1}. The
energy cutoff was set at 500 eV. We allowed spin polarization for
all $TM$B$_4$ but only one compound of MnB$_4$ showed a small
non-zero magnetic moment of about 0.7 $\mu_B$ per Mn in the
antiferromagnetic ordering. A very accurate optimization of
structural parameters was achieved by minimizing forces (below 0.001
eV/\AA) and stress tensors (typically below 0.5 kB). Further
simulation details and the procedure for calculating the mechanical
properties are described in the supplementary material\cite{RefAdd}.


Figure 2 shows our experimental ED patterns projected along the [100],
[110], [111] and [101] directions revealing that the unit cell has the
dimensions $|$a*$|$$\approx$2.1 nm$^{-1}$, $|$b*$|$$\approx$1.8
nm$^{-1}$ , $|$c*$|$$\approx$3.5 nm$^{-1}$ and it can be classified as
a primitive orthorhombic lattice. The simulated ED pattern along [101]
for the \emph{oP}10 structure (Fig.  2f) shows additional reflections
as compared to \emph{oI}10 (Fig.  2d), which is expected because the
unit cell is doubled and the number of symmetry operations is reduced
from 16 to 8. The corresponding extra reflections are clearly present
in the observed [101] pattern which unambiguously points at the
\emph{oP}10 structure. The \emph{oP}10 structural model was further
used to refine the powder XRD data and by that a good agreement
between experiment and theory is obtained (see Table S1 in Ref.
\cite{RefAdd}).

\begin{figure}[t!]
\begin{center}
\includegraphics[width=0.47\textwidth]{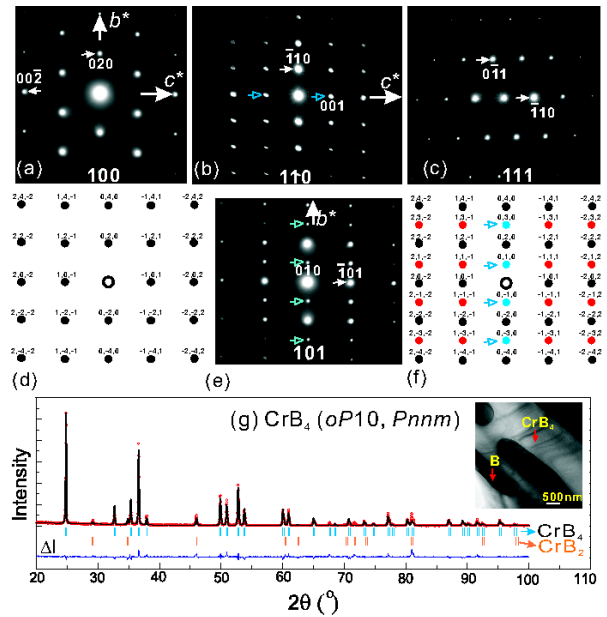}
\end{center}
\caption{(color online) (a-c,e) Experimental electron diffraction
(ED) patterns along the [100], [110], [111] and [101] directions,
respectively. (d,f): theoretical [101] ED patterns for $oI10$ and
$oP10$ structures, respectively; (g) Rietveld refinement (Cu
K$_{\alpha1}$) of the experimental X-ray diffraction data;
reflections of CrB$_2$ are indicated by vertical bars. Hollow arrows
in (b,e,f) denote the second-order diffraction spots. Inset in (g)
is a TEM image showing the phase boundary of CrB$_4$.}\label{fig2}
\end{figure}

The mechanism causing the observed distortion in CrB$_4$ is many
fold, because it is related to the quasi-$sp^3$ B-B bonding, the
hybridization of the B- and Cr-like states, the atomic size of Cr
relative to the available volume in the B cage and the charge
transfer between B and Cr (i.e. the valencies). It is illustrative
to look first at the evolution of the B network in the sequence of
the related structure types: diamond, bct-C$_4$, $oI$10, and $oP$10
(see Fig. 1). In diamond, the tetrahedral arrangement of four
nearest neighbors with the $\cos^{-1}(-1/3)\approx109.5^\circ$
angles between the bonds is optimal for the $sp^3$
hybridization\cite{Pettifor} (Fig. 1(a)). In bct-C$_4$, the symmetry
of the local atomic environment is broken as two bonds form a
90$^\circ$ angle (Fig. 1(b)). In \emph{oI}10, symmetry is further
reduced due to the two bonds now having different lengths (namely,
1.73 \AA\ and 1.86 \AA\ for structurally relaxed CrB$_4$, see Fig.
1(c)). Finally, in \emph{oP}10 a further deviation from the ideal
$sp^3$ geometry occurs as one of the two nonequivalent B sites has
two B-B bonds at an angle well below 90$^\circ$ and the other is no
longer four-fold coordinated (a fifth B atom is 2.13 \AA\ away and
the electron localization function shown in Fig. 1(d) develops a
blob along the short diagonal of the B parallelogram).

Within a semiempirical extended H\"uckel approach Burdett {\it et al.}
\cite{Cr-Mn} studied the relative stability of carbon- and boron-based
structures by analyzing the moments ($\mu_n$) of the electronic
density of states (DOS). They argued, that bct-C$_4$-carbon is less
stable than diamond-carbon because (i) the non-optimal 90$^\circ$
angle gives rise to a strain energy and (ii) the four-membered rings
result in a higher $\mu_4$ which translates into a more unimodal DOS
and a lower stability for elements with a half-filled shell. Their
conclusion that the second contribution could stabilize boron-based
materials with a lower number of electrons is supported by our DFT
calculations: for the group-IV carbon the diamond structure is favored
by 0.20 eV/atom while for the group-III boron the bct-C$_4$ structure
is favored by 0.08 eV/atom. The structural differences within the
bct-C$_4$, \emph{oI}10, and \emph{oP}10 family are less
pronounced. However, the presence of three-membered rings in
\emph{oP}10 may significantly influence the structure's important
third and forth DOS moments.

\begin{figure}[t!]
\begin{center}
\includegraphics[width=0.48\textwidth]{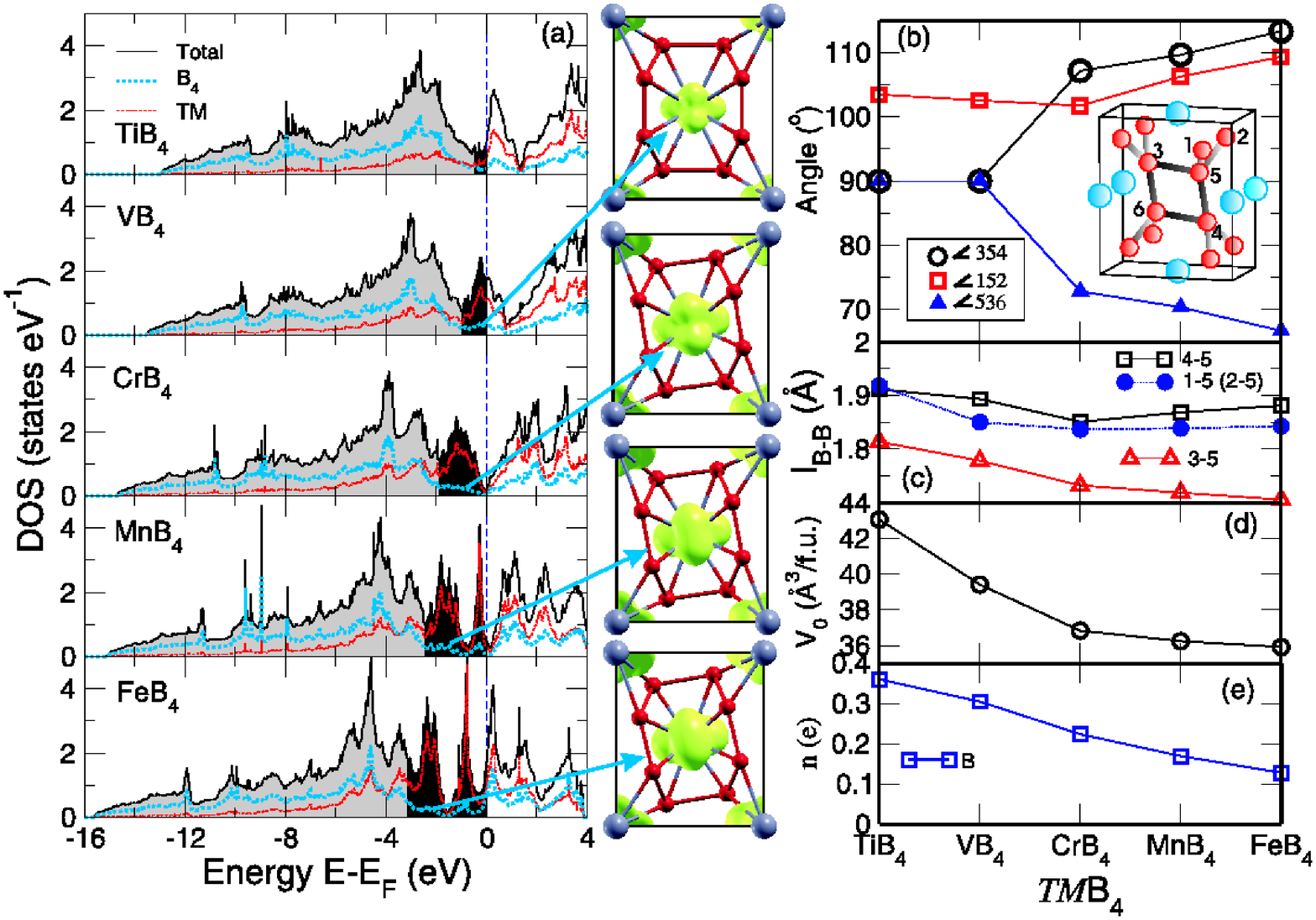}
\end{center}
\caption{(color online) Calculated electronic structure of TMB$_4$
compounds (TM: $3d$ transition elements). (a): electronic densities
of states (DOS) with occupied bonding and nonbonding states
highlighted in grey and black, respectively. The corresponding
insets of structures show charge density isosurfaces (0.1
e/\AA\,$^3$) for the energy windows corresponding to the nonbonding
states. (b-e): variation of bond angles, bond lengths, atomic
volume, and charge transfer to boron obtain by Bader's
decomposition.\cite{Bader}}\label{fig3}
\end{figure}

\begin{figure}[b!]
\begin{center}
\includegraphics[width=86mm,angle=0]{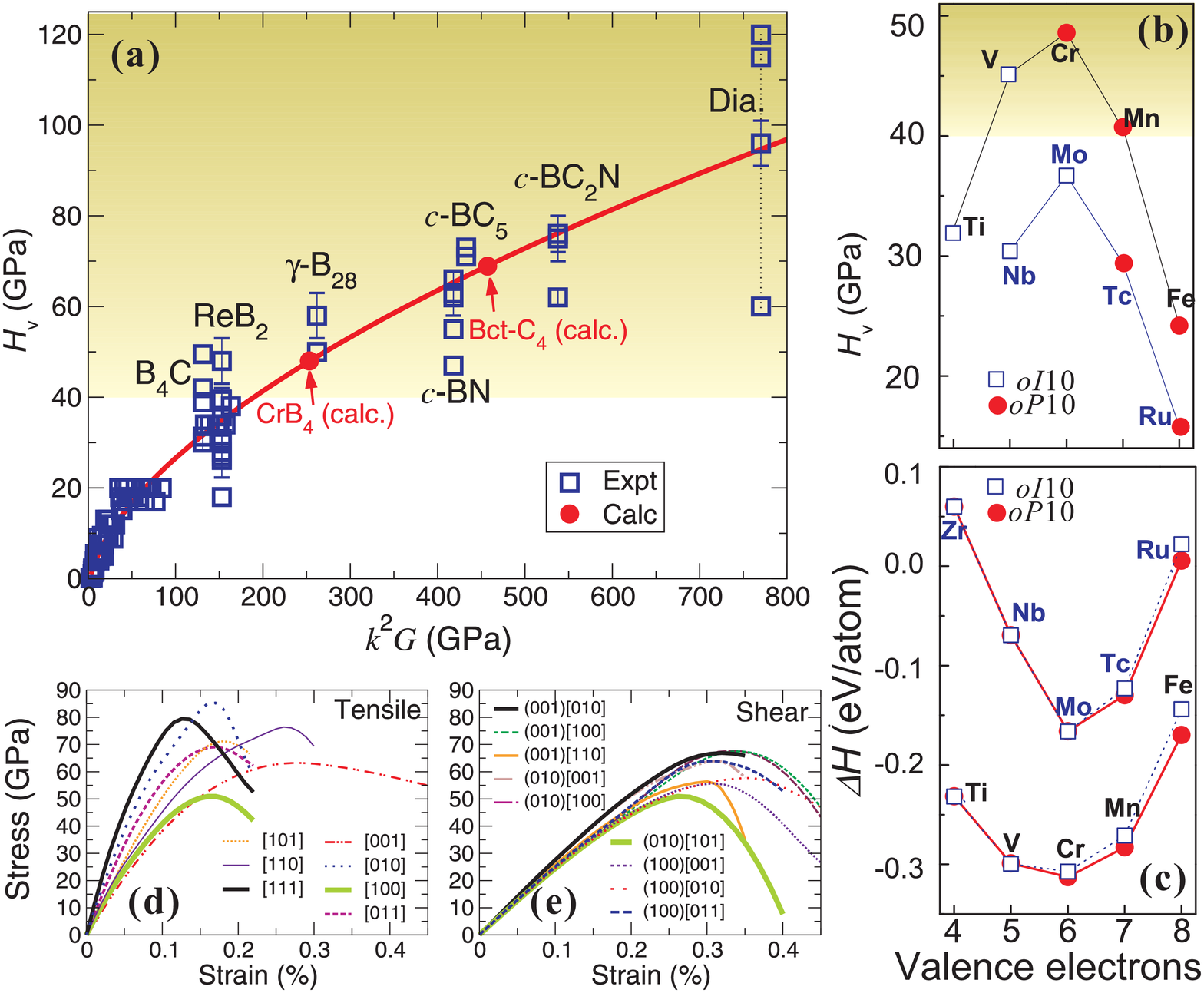}
\end{center}
\caption{(color online) (a): Vickers hardness, $H_v$, as a function of
$k^2G$, with $k$ being the ratio of the shear ($G$) to bulk ($B$)
modulus; the experimental data are discussed in the supplementary
material \cite{RefAdd}; (b): $H_v$ for the most stable TMB$_4$
structures derived from the calculated $B$ and $G$ values (see text
and Ref. \onlinecite{22}); (c): formation enthalpy; (d,e): ideal
tensile and shear strengths of $oP10$-CrB$_4$.} \label{fig4}
\end{figure}

To elucidate the stabilization role of the TM atom we carried out a
series of DFT calculations for ten TMB$_4$ compounds. The DOS and
formation enthalpies are presented for the more stable of the $oI$10
or $oP$10 structure. For the $3d$ series Fig. 3(a) shows rather
similar DOS profiles with the Fermi level ($E_F$) moving upwards as
the electron count increases. In the exemplary CrB$_4$ case, the DOS
in the range of (-14, -5) eV is mostly of B-$s-p$-like character. The
weight of the Cr DOS increases gradually and in the range of (-5,
-2) eV a strong hybridization between Cr-\emph{d}$_{xy}$
(-\emph{d}$_{xz}$) and B-\emph{p}$_y$ (-\emph{p}$_z$) states is
observed. In the region from -2 to 0 eV the
Cr-\emph{d}$_{(x^2-y^2)}$ and Cr-$d_{yz}$ nonbonding states become
dominant. The position of $E_F$ in the pseudogap along with the
lowest formation enthalpy achieved for TM = Cr (see Fig.
\ref{fig4}c) is consistent with the prediction\cite{Cr-Mn} of
maximum stability occurring in the middle of the $3d$ series.
Further DFT calculations as detailed in the supplementary material
\cite{RefAdd} shed light on the stability competition between the
$oP$10 and the $oI$10 structures. We find that Ti and V as well as
Zr, Nb, and Mo tetraborides prefer the $oI$10 type.  Furthermore,
the energy gained by the $oI$10 to $oP$10 transformation is larger
for the $3d$ compounds and increases within both series from left to
right. It is also noticeable, that the structural transformation is
accompanied by a volume reduction whereby over 80\% of the energy
gain comes just from the distortion of the B network. Finally, by
artificially decreasing (increasing) the volume one can induce
(disfavor) the distortion for all the considered TMB$_4$ compounds.
Figure 3(b-e) summarizes the structural trends and shows the
variation of the average Bader's charge \cite{Bader}. We employ the
Bader's charge decomposition to illustrate that the charge transfer
from the TM element to B (which is around 1 e/TM) {\it decreases} in
the sequence from Ti to Fe. Hence, the distortion could be explained
by the decreasing number of electrons transferred to B. The derived
geometrical result, that three out of four B-B bonds have a minimum
bond length for TM = Cr (Fig. 3c) demonstrates further why CrB$_4$
is particularly stable. Considering that $E_F$ in FeB$_4$ moves from
a deep valley in $oI10$ into the shoulder of the antibonding
B-$p$-Fe-$d$ peak in $oP10$ \cite{FeB4}, the optimality of the
$p$-$d$ bonding appears to be of less importance for the compound's
stability (note that the unexpectedly high DOS at $E_F$ in
$oP10$-FeB$_4$ makes the compound a good candidate to be a {\it
phonon-mediated} superconductor with a $T_c$ of 15-20 K
\cite{FeB4}).

\begin{table*}
\caption{Calculated elastic properties (in GPa) for CrB$_4$ and
known (super)hard materials. The calculated bulk (\emph{B}) and
shear moduli (\emph{G}) are Reuss-Voigt-Hill averages. The Vickers
hardness estimates ($H_v^{Calc}$) were obtaned with our proposed
formula \cite{22,22-1} using the calculated elastic moduli. Finally,
the experimental Vickers hardness values ($H_v^{Exp}$) for diamond,
BC$_2$N, bct-C$_4$, c-BN and B$_4$C were taken from  Refs.
\cite{22,Grasso2011,Kislyi1985}. Additional information is given in
Table S5 \cite{RefAdd}.}
\begin{ruledtabular}
\begin{tabular}{ccccccccccccccccccc}
  & \emph{C}$_{11}$ & \emph{C}$_{22}$ &   \emph{C}$_{33}$ &   \emph{C}$_{44}$ &
  \emph{C}$_{55}$  &  \emph{C}$_{66}$ &  \emph{C}$_{12}$ & \emph{C}$_{13}$ & \emph{C}$_{23}$ & \emph{B} &    \emph{G}    &
\emph{H}$_{v}^{Calc}$ & \emph{H}$_{v}^{Exp}$ \\
\hline
Diamond & 1079 &     &     & 578 &     &     & 124 &    &     & 442 & 536 &    95.7       & 96$\pm$5 \\
BC$_2$N &      &     &     &     &     &     &     &    &     & 408   & 445   &    75.4   & 76$\pm$2 \\
bct-C$_4$ & 933 &    & 1190& 447&      &325  &172  & 59 &     & 404   & 421   &     68.9  &  \\
c-BN    & 820  &     &     & 480&      &     & 190 &    &     & 400   & 405   &    65.2   & 66$\pm$2\\
B$_4$C  &     &     &     &     &     &     &     &     &     &247    & 200   &     31.7  &30$\pm$2, 31.3-38.9, 42-49\\
CrB$_4$(oP10) & 554  & 880 &  473&  254&  282&  250&  65 & 107&  95 & 265   & 261  & 48.0 &   \\
CrB$_4$(oI10) & 591  & 931 &  467&  252&  280&  225&  64 & 115&  97 & 275   & 259  & 45.1 &   \\
\end{tabular}
\end{ruledtabular}
\label{tab1}
\end{table*}

The mechanical properties of CrB$_4$ are examined and rationalized via
DFT calculation of the elastic properties for the mentioned ten
TMB$_4$ compounds (see Ref. \cite{RefAdd}). All of them are found to
exhibit ultra incompressibility along the \emph{b}-axis and high bulk
($B$) and sheer ($G$) moduli.  CrB$_4$ is found to have the highest
shear modulus ($G=261$ GPa) and Pugh's ratio \cite{23}
($k=G/B=261/265=0.985$), which are two important elastic properties
thought to be strongly correlated to hardness \cite{22}. The
compound's low Poisson ratio of $v=0.12$ is typical for materials with
strong covalent bonding. \cite{11} Strikingly, the calculated lowest
ideal shear and tensile strengths of 51 GPa are remarkably high and
comparable to the lowest tensile strength of 55 GPa for superhard
\emph{c}-BN \cite{26}. These values exceed considerably the lowest
ideal shear strength of 34 GPa \cite{26} in ReB$_2$ which structure is
comprised of buckled 2D boron nets\cite{15}. As a corroboration for a
possibly outstanding hardness, by breaking Cr-B bonds along the [001]
direction we found the lowest critical cleavage stress \cite{Lazar-1}
of 53 GPa, which matches the lowest ideal strengths. Finally, for
estimation of the Vickers hardness ($H_v$) in terms of elastic
properties we employ a recently proposed empirical
model\cite{22,22-1},
\begin{math} H_v = 2.0(k^2G)^{0.585}-3.0 \end{math}
($H_v$ and $G$ in GPa), which performs well across a large class of
materials and hardness values (Fig. \ref{fig4}, panel (a)). Figure
\ref{fig4} reveals that the predicted behavior of $H_v$ for the
TMB$_4$ compounds (panel b) mirrors the trend in their enthalpy of
formation (panel c). The largest hardness value of $H_v=48$ GPa for
CrB$_4$ is well above the superhardness threshold of $40$ GPa and
decreases rapidly for the considered TMs. In particular, the
isoelectronic but larger Mo atom streches the B network beyond its
optimal size, leading to a 25\% reduction in hardness. When compared
against the known B$_4$C material which can also be synthesized under
ambient pressure, CrB$_4$ displays (see Table I) superior elastic
properties and estimated $H_v$ (note that according to recent
measurements B$_4$C is not superhard in its crystalline form
\cite{Grasso2011}).

Our findings make $oP$10-CrB$_4$ a prime candidate to be an
(up-to-now overlooked) affordable ambient-pressure superhard
material. Measurement of the compound's Vickers hardness will be a
challenge as pure CrB$_4$ samples are difficult to produce with
standard methods due to the particular behaviour of the Cr-B system
in the high-temperature - B-rich part of the binary phase diagram
\cite{cr-b}. Namely, the cooling of an arc-melted 1:4 elemental
mixture leads unavoidably to a two-phase coexistence of CrB$_2$ and
B in a wide high-temperature region from 1830 $^\circ$C to 1500
$^\circ$C. Formation of CrB$_4$ occurs below 1500 $^\circ$C but
significant fractions of CrB$_2$ and B can still be present after
week(s) of sample annealing, as happened in the original \cite{12}
and present \cite{RefAdd} studies. Our currently best samples with
78\% content of CrB$_4$ allowed us to reliably characterize the
compound's crystal structure but were not suitable for investigation
of its mechanical properties. Therefore, alternative approaches,
such as the powder metallurgical process or the single-crystal
growth method, may need to be employed to obtain samples of desired
quality.


The confirmation of the new $oP10$ crystal structure of CrB$_4$ makes
the prospect of synthesizing the FeB$_4$ phase with the same structure
-predicted to be a viable high-temperature and high-pressure ground
state of the Fe-B system \cite{FeB4,14}- more exciting. Our detailed
experimental and theoretical study of the presumably superhard
compound CrB$_4$ demonstrates that materials with appealing properties
may still be found in reportedly well-known binary systems.

{\bf Acknowledgement} We thank Xiaobing Hu and Prof. Shaobo Mi in
the IMR for their valuable helps in performing and analyzing ED
experiments and Prof. Shi Liu for his help in synthesizing
experimental samples. We are grateful for supports from the
``Hundred Talents Project'' of Chinese Academy of Sciences and from
NSFC of China (Grand Numbers: 51074151, 51174188) as well as Beijing
Supercomputing Center of CAS (including its Shenyang branch in the
IMR). A.N.K acknowledges the support from EPSRC CAF EP/G004072/1 in
the UK.


\begin{thebibliography}{99}

\bibitem{1} R.B. Kaner,
            J.J. Gilman,
        and S.H. Tolbert,
            Science, \textbf{208}, 1268 (2005).

\bibitem{1-1}  V. V. Brazhkin,
               A. G. Lyapin,
           and R, J. Hemley,
               Philo. Mag. A, \textbf{82}, 231 (2002).

\bibitem{1-2} V. V. Brazhkin,
              N. Dubrovinskaia,
              M. Nicol,
              N. Novikov,
              R. Riedel,
              V. Solozhenko,
          and Y. Zhao,
              Nat Mater, \textbf{3}, 576 (2004).

\bibitem{1-3} J. J. Gilman. Science \textbf{261}, 143 (1993).

\bibitem{1-4} S. Veprek, J. Nanosci. Nanotechnol., \textbf{11}, 14 (2011).

\bibitem{2} V. L. Solozhenko,
            O. O. Kurakevych,
            D. Andrault,
            Y. L. Godec,
        and M. Mezouar,
            Phys. Rev. Lett., \textbf{102}, 015506 (2009).

\bibitem{3} V. L. Solozhenko,
            O. O. Kurakevych
        and A. R. Oganov,
            J. Superhard Mater., \textbf{30}, 428 (2008).

\bibitem{3-1} A. R. Oganov,
              J. H. Chen,
              C. Gatti,
              Y. Z. Ma,
              Y. M. Ma,
              C. W. Glass,
              Z. X. Liu,
              T. Yu,
              O. O. Kurakevych,
          and V. L. Solozhenko,
              Nature, \textbf{457}, 863 (2009).

\bibitem{3-2} E. Y. Zarechnaya,
              L. Dubrovinsky,
              N. Dubrovinskaia,
              Y. Filinchuk,
              D. Chernyshov,
              V. Dmitriev,
              N. Miyajima,
              A. El Goresy,
              H. F. Braun,
              S. Van Smaalen,
              I. Kantor,
              A. Kantor,
              V. Prakapenka,
              M. Hanfland,
              A. S. Mikhaylushkin,
              I. A. Abrikosov,
          and S. I. Simak,
              Phys. Rev. Lett., \textbf{102}, 185501 (2009).

\bibitem{4} Q. Li,
            Y. Ma,
            A. R. Oganov,
            H. Wang,
            H. Wang,
            Y. Xu,
            T. Cui,
            H. K. Mao,
        and G. Zou,
            Phys. Rev. Lett. \textbf{102}, 175506 (2009).

\bibitem{5} K. Umemoto,
            R. M. Wentzcovitch,
            S. Saito,
        and T. Miyake,
            Phys. Rev. Lett. \textbf{104}, 125504 (2010).

\bibitem{5-1} R. H. Baughman
          and D. S. Galvao,
              Phys. Rev. Lett. \textbf{104}, 125504 (2010).

\bibitem{6} J.-T. Wang,
            C. Chen,
        and Y. Kawazoe,
            Phys. Rev. Lett. \textbf{106}, 075501 (2011).

\bibitem{7} W.L. Mao,
            H.-k. Mao,
            P. J. Eng,
            T.P. Trainor,
            M. Newville,
            C.-c. Kao,
            D. L. Heinz,
            J. F. Shu,
            Y. Meng
        and R. J. Hemley,
            Science, \textbf{302}, 425 (2003).

\bibitem{7-2} H. Y. Niu, 
              X.-Q. Chen,
              S. B. Wang,
              D. Z. Li,
              W. L. Mao,
              Y. Y. Li,
              Phys. Rev. Lett., \textbf{108}, 135501 (2012).

\bibitem{8} P. Y. Wei,
            Y. Sun,
            X.-Q. Chen,
            D. Z. Li,
        and Y. Y. Li,
            Appl. Phys. Lett., \textbf{97}, 061910 (2010).

\bibitem{8-1} F.M. Gao
          and X.F. Hao,
              Phys. Stat. Sol. (RRL) - Rapid Res. Let., \textbf{4},
              200 (2010).

\bibitem{8-2} H.Y. Niu,
              P.Y. Wei,
              Y. Sun,
              X.-Q. Chen,
              C. Franchini.,
              D.Z. Li,
          and Y.Y. Li,
              Appl. Phys. Lett., \textbf{99}, 031901 (2011).

\bibitem{8-3} Q. Zhu,
              A. R. Oganov,
              M. A. Salvad\'o,
              P. Pertierra,
          and A. O. Lyakhov,
              Phys. Rev. B, \textbf{83}, 193410 (2011).

\bibitem{11} J. B. Levine,
             S. H. Tolbert,
         and R. B. Kaner,
             Adv. Funct. Mater, \textbf{19}, 3519 (2009).

\bibitem{Mcmillan} P. F. Mcmillan,
                   Nat. Mat., \textbf{1}, 19 (2002).

\bibitem{ReB2} J.B. Levine,
               J.B. Betts,
               J.D. Garrett,
               S.Q. Guo,
               J.T. Eng,
               A. Migliori,
           and R.B. Kaner,
               Acta Mater, \textbf{58}, 1530 (2010).

\bibitem{ReB2-1} H.-Y. Chung,
                 M.B. Weinberger,
                 J.B. Levine,
                 A. Kavner,
                 J.-M. Yang
             and S.H. Tolbert,
                 Science \textbf{316}, 436 (2007).

\bibitem{ReB2-2} J.B. Levine,
                 S.L. Nguyen,
                 H.I. Rasool,
                 J.A. Wright,
                 S.E. Brown
             and R.B. Kaner,
                 J. Am. Chem. Soc. \textbf{130}, 16953 (2008).


\bibitem{FeB4} A. N. Kolmogorov,
               S. Shah,
               E. R. Margine,
               A. F. Bialon,
               T. Hammerschmidt,
           and R. Drautz,
               Phys. Rev. Lett.,
               \textbf{105}, 217003 (2010).

\bibitem{14}   A. F. Bialon,
               T. Hammerschmidt,
               R. Drautz,
               S. Shah,
               E. R. Margine,
           and A. N. Kolmogorov,
               Appl. Phys. Lett., \textbf{98}, 081901 (2011).




\bibitem{PBE}
             J.P. Perdew,
             K. Burke,
         and M. Ernzerhof,
             Phys. Rev. Lett. {\bf 77}, 3865 (1996).

\bibitem{PAW}
             P. E. Bl\"{o}chl,
             Phys. Rev. B, {\bf 50}, 17953 (1994).

\bibitem{VASP}
             G. Kresse
         and J. Hafner,
             Phys. Rev. B {\bf 47}, 558 (1993).

\bibitem{VASP-1} G. Kresse
             and J. Furthm\"{u}ller,
                 Phys. Rev. B {\bf 54}, 11169 (1996).

\bibitem{22} X.-Q. Chen,
             H. Y. Niu,
             D. Z. Li
         and Y. Y. Li,
             Intermetallics, \textbf{19}, 1275 (2011).

\bibitem{22-1} X.-Q. Chen,
               H. Y. Niu,
               C. Franchini,
               D. Z. Li
           and Y. Y. Li,
               Phys. Rev. B, \textbf{84}, 121405(R) (2011).

\bibitem{12} S. Andersson,
         and T. Lundstr\"om,
             Acta Chem. Scand., \textbf{22}, 3103 (1968).

\bibitem{13} H. B. Xu,
             Y. X. Wang
         and V. C. Lo,
             Phys. Stat. Sol. (RRL) - Rapid Res. Let., \textbf{5}, 13 (2011).


\bibitem{27} B. Silvi
         and A. Savin,
             Nature (London), \textbf{371}, 683 (1994).

\bibitem{RefAdd} See supplementary material at
                 http://link.aps.orgsupplementary/.

\bibitem{fullprof} J. Rodr\"uez-Carvajal, Physica B, \textbf{192}, 55 (1993).







\bibitem{Pettifor} D. G. Pettifor,
                   Bonding and Structure of Molecules and Solids,
                   (Clarendon Press) 1995.

\bibitem{Cr-Mn} J. K. Burdett
            and S. Lee,
                J. Am. Chem. Soc. {\bf 107}, 3063 (1985).


\bibitem{Bader} W. Tang,
                E. Sanville
            and G. Henkelman,
                J. Phys.: Condens. Matter \textbf{21}, 084204 (2009).


\bibitem{23} S. F. Pugh,
             Philos. Mag. Ser. 7, \textbf{45}, 823 (1954).

\bibitem{26} R. F. Zhang,
             S. Veprek,
         and A. S. Argon,
             Appl. Phys. Lett.,
             \textbf{91}, 201914 (2007).

\bibitem{15} X.-Q. Chen,
             C. L. Fu,
             M. Kr\v{c}mar,
         and G. S. Painter,
             Phys. Rev. Lett., \textbf{100}, 196403 (2008).

\bibitem{Lazar-1} P. Lazar
              and R. Podloucky,
                  Phys. Rev. B, \textbf{78}, 104114 (2008).

\bibitem{Grasso2011} S. Grasso,
                     C. Hu,
                     O. Vasylkiv,
                     T. S. Suzuki,
                     S. Guo,
                     T. Nishimura
                 and Y. Sakka,
                     Scripta Materialia, \textbf{64}, 256 (2011).

\bibitem{cr-b} T. B. Massalski
           and H. Okamoto,
               Binary Alloy Phase Diagrams,
               (ASM Intl; 2nd ed., ASM International 1990).


\bibitem{Kislyi1985}
                  P. S. Kislyi,
                  Superhard and Refractory Materials,
                  (Kiev: Institute of Superhard Materials), p.86, 1985





\end{thebibliography}
\end{document}